# Observation of spin-valley locked nodal lines in a quasi-2D altermagnet

Quanxin Hu[1, *], Xingkai Cheng[2, *], Qingchen Duan[1, *], Yudong Hu[1, *], Bei Jiang[1], Yusen Xiao[1], Yaqi Li[1], Mojun Pan[1], Liwei Deng[3], Changchao Liu[4], Guanghan Cao[4, 5, 6], Zhengtai Liu[7], Mao Ye[7], Shan Qiao[3], Zhanfeng Liu[8], Zhe Sun[8, 9, 10, 11], Anyuan Gao[1], Yaobo Huang[7, †,] Ruidan Zhong[1, †], Junwei Liu[2, †], Baiqing Lv[1, 12, †], Hong Ding[1, 13, 14]

[1] Tsung-Dao Lee Institute and School of Physics and Astronomy, Shanghai Jiao Tong University, Shanghai 201210, China

[2] Department of Physics, The Hong Kong University of Science and Technology, Hong Kong, China

[3] Shanghai Institute of Microsystem and Information Technology, Chinese Academy of Sciences, Shanghai, China

[4] School of Physics, Zhejiang University, Hangzhou, China

[5] Institute of Fundamental and Transdisciplinary Research, and State Key Laboratory of Silicon and Advanced Semiconductor Materials, Zhejiang University, Hangzhou, China

[6] Collaborative Innovation Centre of Advanced Microstructures, Nanjing University, Nanjing, China

[7] Shanghai Synchrotron Radiation Facility, Shanghai Advanced Research Institute, Chinese Academy of Sciences, Shanghai 201204, China

[8] National Synchrotron Radiation Laboratory, University of Science and Technology of China, Hefei, Anhui 230029, China

[9] Department of Physics, CAS Key Laboratory of Strongly Coupled Quantum Matter Physics, University of Science and Technology of China, Hefei, Anhui 230026, China

[10] Zhejiang Institute of Photoelectronics, Jinhua, Zhejiang 321004, China

[11] Collaborative Innovation Center of Advanced Microstructures, Nanjing, Jiangsu 210093, China

[12] Zhangjiang Institute for Advanced Study, Shanghai Jiao Tong University, Shanghai 200240, China

[13] Hefei National Laboratory, Hefei 230088, China

[14] New Cornerstone Science Laboratory, Shanghai 201210, China

*These authors contributed equally to this work

†Correspondence to: huangyaobo@sari.ac.cn; rzhong@sjtu.edu.cn; liuj@ust.hk; baiqing@sjtu.edu.cn

**Abstract**

The interplay among quantum degrees of freedom—spin, orbital and momentum—has emerged as a fertile ground for realizing magnetic quantum states with transformative potential for electronic and spintronic technologies. Prominent examples include ferromagnetic Weyl semimetals and antiferromagnetic axion insulators. Recently, altermagnets have been identified as a distinct spin-splitting class of collinear antiferromagnets, characterized by crystal symmetry (rotation/mirror) that connects magnetic sublattices in real space and enforces $C$-paired spin-momentum locking in reciprocal space. These materials combine the advantages of nonrelativistic spin-polarization akin to ferromagnets and vanished net-magnetization as antiferromagnets, making them highly promising for spintronic applications. Furthermore, they introduce nontrivial spin-momentum locking spin texture as an additional degree of freedom for realizing novel quantum phases. In this work, we report the discovery of a new type of spin–valley–locked nodal line phase in the layered altermagnet Rb-intercalated $V_2Te_2O$. By combining high-resolution spin and angle-resolved photoemission spectroscopy with first-principles calculations, we observe the coexistence of both spinless and spinful nodal lines near the Fermi level. Remarkably, the spinful nodal lines exhibit uniform spin polarization within each valley, while displaying opposite spin polarizations across symmetry-paired valleys—a unique feature we term spin–valley–locked nodal lines, which is exclusive to altermagnets. Direct measurements of out-of-plane band dispersion using a side-cleaving technique reveal the two-dimensional nature of these nodal lines, enabling potential manipulation via heterostructure engineering, electrostatic gating, strain, and more. Our findings not only unveil a previously unexplored topological phase in altermagnets where valley-locked spin as an additional quantum character but also establish $RbV_2Te_2O$ as a promising platform for spintronics, valleytronics, and moiré-engineered quantum devices.

**I. Introduction**

Magnetism, one of the oldest studied physical phenomena, continues to play a pivotal role in condensed-matter physics. Traditionally, magnetic materials are classified by their real-space magnetic order, such as ferromagnets (FMs) and antiferromagnets (AFMs), which exhibit nonzero and vanishing net magnetization, respectively. When magnetic order couples to other degrees of freedom, such as orbital character or crystal momentum, a variety of emergent quantum phases can appear [Fig. 1], including but not limited to quantized anomalous Hall effect [1], ferromagnetic Weyl semimetals [2,3] and antiferromagnetic axion insulators [4,5]. These states possess distinctive physical properties and hold great promise for electronic and spintronic applications, making the identification and manipulation of diverse magnetic quantum states a central frontier in modern condensed-matter physics.

Nodal-line semimetals, viewed as an extension of Weyl and Dirac semimetals, host one-dimensional band crossings that form continuous loops or lines in momentum space. This unique geometry gives rise to a range of unconventional physical phenomena, most notably 'drumhead' surface states with a high density of states that amplify correlation effects, such as surface superconductivity, magnetism, and charge order [6,7]. Experimentally and theoretically, nodal lines have been identified in various magnetic systems. For example, AFMs can host fourfold-degenerate Dirac nodal line formed by spin-degenerate bands [8], while FMs—due to Zeeman splitting—typically exhibit twofold-degenerate nodal lines composed of majority-spin bands near the Fermi level, giving rise to spin-polarized Weyl nodal lines [9] [Fig. 1]. However, in all such magnetic systems reported to date— whether in spin-degenerate AFMs and uniformly spin-polarized FMs—spin has not been utilized as a binary quantum index to characterize the nodal lines, nor has it been coupled with additional internal degrees of freedom such as orbital or valley channels. This limits both the tunability of their topological properties and the diversity of emergent quantum phenomena they can support, thus motivating research for magnetic systems in which spin serve as a tunable index.

Altermagnets (AMs) represent a special type of collinear AFMs with nonrelativistic spin-splitting, characterized by specific crystal symmetry $C$ (mirror or rotation), which connects opposite-spin sublattices in real space while leads to $C$-paired spin-valley locking (SVL) in reciprocal space [10–16], and described by spin group symmetry [17–19]. AMs combine nonrelativistic spin splitting, as in FMs, with a vanishing net magnetization, as in AFMs, enabling spin-polarized states for information storage while remaining robust against external perturbations. Several altermagnetic candidates have now been experimentally confirmed.

Photoemission spectroscopy has provided compelling evidence for the altermagnetic phase in materials such as $\alpha$-MnTe [20], CrSb [21], MnTe$_2$ [22], CoNb$_4$Se$_8$ [23–25], (Rb, Cs)V$_2$Te$_2$O [26–28]. and KV$_2$Se$_2$O [26,27], and only the latter two are only d-wave altermagnet that can support nonrelativistic spin current. Owing to the $C$-paired nature in both real and reciprocal space, AMs exhibit a range of unconventional physical phenomena— including spin splitting torque [10,29–31], giant and tunneling magnetoresistance [32], piezomagnetic effect [16], anomalous Hall effect [33–37], higher-order topological states [38] and altermagnetic ferroelectricity [39], which prompts AMs as promising candidates for spintronic, valleytronics, twistronics applications [40–43] and superconducting heterostructures [44–52]. More importantly, the coexistence of spin, orbital, and valley degrees of freedom in AMs opens new avenues for realizing unconventional, symmetry-protected nodal-line structures that remain largely unexplored.

In this study, by combining high-resolution spin and angle-resolved photoemission spectroscopy with first-principles calculations for layered AM RbV$_2$Te$_2$O, we identify two distinct types of nodal lines in the absence of spin–orbit coupling (SOC) due to its dual combined features of both FMs and conventional AFMs: (i) spinful nodal lines, arising from band crossings between states with the same spin within a single valley but exhibiting opposite spin polarization across symmetry-paired valleys — a phenomenon we term spin–valley–locked nodal lines (SVLNLs), which are robust against SOC; and (ii) spinless nodal lines, originating from crossings between bands with opposite spin, which can be gapped by SOC. Furthermore, by utilizing side-surface cleaving techniques, we directly probe the band dispersion along the out-of-plane ($k_z$) direction, revealing negligible interlayer coupling in RbV$_2$Te$_2$O, thereby classifying it as a quasi-two-dimensional (quasi-2D) AM. Our findings not only reveal the previously unobserved topological phases in AMs but also prompt RbV$_2$Te$_2$O as an ideal platform for artificial structuring, opening new avenues for spintronic and valleytronics applications.

## II. Results

RbV$_2$Te$_2$O crystallizes in a van der Waals-like tetragonal layered structure with space group $P4/mmm$ (No. 123). Its structure comprises V$_2$Te$_2$O trilayers separated by Rb atomic layers [Fig. 2(a) and Fig. 8]. The Rb atomic layers in RbV$_2$Te$_2$O primarily serve as electron donors to the V$_2$Te$_2$O layers, with minimal impact on the electronic band structure. The variation of cleaving temperature modifies the residual Rb coverage on the cleaved surface, acting as an extrinsic surface-doping effect that merely shifts the Fermi level, without altering the intrinsic band topology (see Fig. 9). Within each trilayer, the $V_2O$ plane forms a variant of the Lieb lattice, where the antiparallel magnetic moments form two distinct sublattices ($V_A$ and $V_B$) related by fourfold rotation $C_4$ and mirror $M_{xy}(M_{\bar{x}y})$, which is sandwiched between two Te layers. The $V_A$ and $V_B$ sublattices experience distinct local environments due to the presence of the nonmagnetic O site, thereby giving rise to different next-neighbor hopping terms $t_{2a}$ and $t_{2b}$ [Fig. 2(b)]. Figure 2(c) displays the experimentally measured Fermi surface (FS) at $T$ = 20 K, exhibiting four-fold rotational symmetry in the $k_x$–$k_y$ plane. The corresponding first-principles calculation [Fig. 2(d)] shows excellent agreement with the ARPES measurement. The FS is dominated by three bands, $α$, $β$, and $γ$. Among them, the $α$ and $γ$ bands carry the same spin polarization, while the $β$ band exhibits the opposite sign. Moreover, the spin polarization reverses between the X–M and Y–M directions, revealing an alternating $d$-wave spin texture [9].

To verify the spin polarization of the band structure, we performed spin- and angle-resolved photoemission spectroscopy (spin-ARPES). We selected two symmetry-related spectral cuts — cut 1 along Y–M and cut 2 along X–M [Fig. 2(d)] — to directly demonstrate the $C$-paired SVL spin texture. ARPES intensity plots along both cuts resolve three well defined bands $α$ ($α'$), $β$ ($β'$), and $γ$ ($γ'$). Then, we measured the spin polarizations of the ARPES spectra near $E_F$ along these two cuts. Figure 3(a) and (b) show that the spin polarizations have down-up-down and up-down-up sequences along cuts 1 and 2, respectively, in agreement with the calculations. It should be noted that the maximum spin polarization observed in the spectra is about 10%. It is noteworthy that the maximum spin polarization observed in the spectra reaches only about 10%. Although this value aligns with previous spin-ARPES reports [20,22,26,27], it remains significantly below theoretical expectations. A plausible origin of this discrepancy lies in the presence of multiple antiferromagnetic domains, where the spin-ARPES signal effectively averages over differently oriented domains, thereby suppressing the net spin polarization detected experimentally. This observation highlights the necessity of domain engineering or the use of nano-focused probes to directly access the intrinsic spin polarization characteristic of altermagnets.

Building on the identification of $RbV_2Te_2O$ as a $d$-wave spin-momentum locking altermagnet, we conducted a more in-depth investigation of its electronic band structure. The ARPES spectra shown in Fig. 4(a) were measured along high-symmetry direction at 20K using $hv$ = 70eV linearly horizontal (LH) polarized light. The calculated band structure aligns well with the measured band dispersions, which have identical band dispersion along Γ–X–M and Γ–Y–M direction while reversed spin polarization due to $C$-paired SVL. They originate from different sublattices and exhibit opposite spin-splitting between symmetry-connected X and Y valleys [Fig. 4(a)]. The momentum-dependent spin polarization drives a phase with two majority spin bands near the Fermi level and gives rise to multiple twofold degeneracies on the mirror planes $k_z = 0$. These degeneracies, arising from crossings between bands with either identical or opposite spin character, give rise to two distinct types of nodal lines: spinful nodal lines formed by crossings of same-spin bands with opposite spin polarizations in different valleys, which we term spin-valley locked nodal lines (SVLNLs) and spinless nodal lines formed by opposite-spin crossings [Fig. 4(b)]. These nodal lines are protected by crystal symmetry. In the absence of SOC, where spin-up and spin-down channels are decoupled, these nodal lines emerge from the different orbital characteristics of V atoms. Specifically, bands around the fermi level are contributed by the $d_{xz}^\uparrow(M_z = -1)$ and $d_{xy}^\uparrow(M_z = +1)$ orbitals of the $V_A$ atom, $d_{yz}^\downarrow(M_z = -1)$ and $d_{xy}^\downarrow(M_z = +1)$ from the $V_B$, as shown in Fig. 4(c), where $M_z = +1(-1)$ is plotted as solid (dashed) circle. The SVLNLs are protected by $M_z$ symmetry, where two bands with opposite eigenvalues of $M_z$ are disallowed to hybridize with each other, while spinless crossings occur between decoupled spin channels [Fig. 4(c)]. Introducing SOC hybridizes spins but preserves $M_z = \pm i$ eigenvalues, gapping spinless nodal lines while leaving SVLNLs intact (see Fig. 10)). This demonstrates the interplay between CSVL and symmetry-protected topology in altermagnets.

According to our theoretical analysis, two distinct classes of nodal lines—spin–valley–locked nodal lines (SVLNLs) and spinless nodal lines—are expected to reside around the X and Y pockets [Fig. 5(a)]. To resolve these features experimentally, we perform energy–momentum ($E$–$k_x$) cuts along directions intersecting the nodal line structures, as indicated by the cyan and red lines in Fig. 5(a). These cuts reveal multiple band crossings that persist over a finite range of $k_y$ [Figs. 5(b) and 5(c)], indicative of a series of nodal points formed by the touching of upper and lower bands. The band crossings can be quantitatively fitted using multiple Lorentzian peaks, with crossing points highlighted by cyan and orange four-pointed stars. Guided by spin-resolved ARPES data [Fig. 3] and DFT results [Fig. 4], we attribute these crossings to bands derived from distinct orbitals with differing spin characters. Crossings between bands of identical spin polarization correspond to SVLNLs (cyan stars), whereas those between bands of opposite spin signify spinless nodal lines (orange stars). Notably, these nodal points are confined within a narrow energy window [Figs. 5(b) and 5(c)], enabling their systematic tracking across constant-energy surfaces at different $E_B$. Figures 5(d) and 5(e) show constant-energy contours at $E_B$ = 0.2 eV and 0.4 eV, respectively, revealing the evolution of the nodal-line structures. The positions of spinless nodal lines and SVLNLs are marked by color stars, directly extracted from the $E$–$k_x$ slices. These results provide compelling experimental evidence for the coexistence of spinless and spin–valley–locked nodal lines in RbV$_2$Te$_2$O.

Altermagnets serve as a versatile playground to explore the rich interplay of magnetism with multiple degrees of freedom—orbital, momentum, and valley. To fully harness the potential of altermagnets in artificial structuring, it is essential to demonstrate that the interlayer coupling in RbV$_2$Te$_2$O is sufficiently weak, thereby ensuring the preservation of its intrinsic altermagnetic properties even in the two-dimensional limit. To investigate the interlayer coupling strength in the altermagnet RbV$_2$Te$_2$O, accessing the side a–c surface is essential. We successfully cleaved this surface using a top post and silver epoxy. Synchrotron-based ARPES measurements were then performed to obtain real-space maps of photoemission intensity on the cleaved samples [Fig. 6(a)], allowing us to identify well-cleaved regions with homogeneous intensity. ARPES measurements clearly revealed the electronic band structure on the side a–c surface. Utilizing $LH$ polarized incident light with photon energies of $hv$ = 81 eV and $hv$ = 111 eV—corresponding to the $k_y$ = 0 and $k_y$ = $\pi$ planes (see Fig. 11), respectively [Fig. 6(b),]—we measured the FSs and band dispersions. As shown in Figs. 6(c) and 6(d), the FSs in both the $k_y$ = 0 and $k_y$ = $\pi$ planes, exhibit quasi-one-dimensional characteristics, with negligible band dispersion along the $k_z$ direction (Fig. 12). Figures 6(c) and 6(d) present three-dimensional representations of the band structure in the $k_y$ = 0 and $k_y$ = $\pi$ planes, respectively. The bands display nearly identical dispersions along Γ–X–Γ (M-Y-M) and R-Z-R (A-T-A) directions (Fig. 13). Our results demonstrate unambiguously that RbV$_2$Te$_2$O is a quasi-2D altermagnet with negligible interlayer coupling. This makes RbV$_2$Te$_2$O a promising platform to study the altermagnetism in the two-dimensional limit.

**III. Discussion**

Above we not only provided compelling evidence for the existence of two types nodal lines in altermagnet RbV$_2$Te$_2$O, but also demonstrate that RbV$_2$Te$_2$O is an ideal quasi-2D altermagnet with negligible interlayer coupling. Owing to CSVL, the SVLNLs arise from band crossings that retain the same spin character within an individual valley, yet exhibit opposite spin polarizations across the symmetry-paired X and Y valleys. These valley-locked spin-polarized nodal lines are fundamentally distinct from those found in conventional systems. The SVLNLs are protected by $M_z$ symmetry, which persists even in the presence of spin–orbit coupling: the constituent bands retain their crossings without opening a gap or splitting into Weyl points, in stark contrast to the behavior observed in many other magnetic topological materials [53].

Beyond charge and spin, CSVL also establish valley as a tunable degree of freedom. In recent years, utilizing this valley degree of freedom to store and manipulate information has inspired the emerging field of valleytronics, opening new avenues for electronic and quantum applications [54]. Many crystalline systems host potentially useful electronic valleys, particularly in hexagonal two-dimensional (2D) materials such as graphene and monolayer group-VI transition metal dichalcogenides (TMDs)—including $MoS_2$, $MoSe_2$, $WS_2$ and $WSe_2$—where the low-energy band edges are governed by the time-reversal symmetry paired ±$K$ valleys located at the Brillouin zone corners. In these TMDs, spin–valley locking is dictated by time-reversal symmetry $T$, which connects valleys with opposite spin polarization, and typically necessitates both strong spin–orbit coupling (SOC) and broken inversion symmetry to lift the spin degeneracy. Altermagnets, by contrast, offer a distinct symmetry route to valleytronics. Rather than relying on $T$, they employ crystalline symmetries—such as four-fold rotational ($C_4$) or mirror symmetries—to generate momentum-locked spin polarization even in the absence of SOC. In this context, altermagnets emerge as a highly tunable platform for valley-based control of quantum states. Particularly, the negligible interlayer coupling in $RbV_2Te_2O$ enables its classification as a quasi-2D altermagnet. This opens the door to artificial structuring—including layer stacking, twist-angle modulation, and heterostructure engineering—providing fertile ground for the realization and manipulation of exotic quantum phases rooted in spin, valley, superconductivity and topology (Fig. 4d).

In summary, combining high-resolution spin and angle-resolved photoemission spectroscopy with DFT calculations, we have conducted a comprehensive study of the metallic altermagnet $RbV_2Te_2O$. We observed the coexistence of two types of nodal lines in the absence of spin-orbit coupling (SOC): spin-valley–locked nodal lines and spinless nodal lines. Strikingly, the spin-valley–locked nodal lines in $RbV_2Te_2O$ remain intact despite the presence of spin–orbit coupling (SOC), in contrast to conventional magnetic topological systems where SOC typically lifts nodal-line degeneracies, yielding pairs of Weyl points with opposite chirality. Leveraging side-cleaving approach, we directly access the out-of-plane electronic structure and uncover negligible interlayer coupling, thereby establishing $RbV_2Te_2O$ as a quasi-two-dimensional (quasi-2D) altermagnet. This two-dimensional character, combined with its novel topological phases, establishes RbV$_2$Te$_2$O as a promising platform for artificial structuring, paving the way for new directions in spintronics, valleytronics, and moiré-engineered quantum devices.

## IV. Methods

### A. Single-crystal synthesis and sample characterization

Single crystals of RbV$_2$Te$_2$O were synthesized using Rb-Te flux. The Rb : V : Te : O molar ratio of the initial mixture was 6.3 : 2 : 7 : 1. The reactants were loaded in an alumina crucible after grounding, and then sealed in an evacuated quartz ampule. After pre-reaction at 100 °C for 24h and regrinding in an Ar-filled glove box, the powders were sealed in a stainless steel tube under argon, which was subsequently vacuum-sealed in an evacuated quartz ampule. The assembly was gradually heated to 950°C for over 15h in a muffle furnace, held there for 10 h, and then gradually cooled to 650°C at a rate of 1°C/ h. The black, air-sensitive crystals are separated from flux by centrifugation. X-ray diffraction analysis was performed using a Bruker D8 AdvanceEco diffractometer, collecting data from 10- 90 ° in 2$\theta$ at room temperature. The diffraction patterns can be well indexed by the (*00L*) reflections. DC magnetization measurements down to 2K were executed on a VSM integrated with a Physical Property Measurement System (PPMS).

### B. ARPES measurement

ARPES measurements were performed at BL-09U and BL-03U of the Shanghai Synchrotron Radiation Facility (SSRF), all equipped with a SCIENTA Omicron DA30L analyzer. The energy resolution at BL-09U and BL-03U (SSRF) was 10–60 meV, with an angular resolution of 0.1°. Samples were cleaved in situ and measured under a base vacuum better than 5 × 10−11 Torr. Spin-resolved ARPES measurements were performed using a Scienta Omicron DA30L electron analyzer integrated with a very-low-energy diffraction (VLEED) spin detector at beamline 13U of the NSRL. The measurements were conducted with LH-polarized scattering geometry, with h$\nu$ = 23 eV. The angle resolution was better than 0.1°, and the combined instrumental energy resolution was better than 25 meV.

### C. DFT calculation

The calculations were performed in the framework of density functional theory as implemented in Vienna ab initio simulation package [55]. The projector-augmented wave potential was adopted with the plane-wave energy cutoff set at 440 eV. The exchange-correlation functional of the Perdew–Burke–Ernzerhof type has been used for both structural relaxations and self-consistent electronic calculations, with the convergence criteria as $10^{-5}$ eV [55,56]. The GGA+$U$ method was employed to treat the strong correlations of the V 3$d$ orbitals, where the value of the Hubbard $U$ was taken as 1 eV, which provided the best fitting to the ARPES results. The Brillouin zone was sampled by a $11 \times 11 \times 5$ Γ-centered Monkhorst–Pack mesh. We construct Wannier tight-binding model Hamilton by the WANNIER90 interface [57], with V 3$d$ orbitals and Te 5$p$ orbitals, the nodal lines and fermi surfaces are calculated by Wannier model with $201 \times 201 \times 1$ k-mesh.

**Acknowledgements**

We thank the Shanghai Synchrotron Radiation Facility (SSRF) of BL09U (31124.02.SSRF.BL09U) and BL03U (31124.02.SSRF.BL03U) for the assistance on ARPES measurements. B.L. acknowledges support from the Ministry of Science and Technology of China (Grant No. 2023YFA1407400), the National Natural Science Foundation of China (Grant No. 12374063), the Shanghai Natural Science Fund for Original Exploration Program (Grant No. 23ZR1479900), and the Cultivation Project of Shanghai Research Center for Quantum Sciences (Grant No. LZPY2024). J. L. acknowledges support from the Ministry of Science and Technology of China (Grant No. 2021YFA1401500), Hong Kong Research Grants Council (Grant No. 16303821, 16306722, 16304523 and C6046-24G). R.Z. acknowledges support from the Ministry of Science and Technology of China (Grant No. 2022YFA1402702, 2021YFA1401600), National Natural Science Foundation of China (Grants No. 12334008, and No. 12374148). Y. B. H. acknowledges support by the Shanghai Committee of Science and Technology (Grant No. 23JC1403300), and the Shanghai Municipal Science and Technology Major Project. H. D. acknowledges support from the New Cornerstone Science Foundation (Grant No. 23H010801236), Innovation Program for Quantum Science and Technology (Grant No. 2021ZD0302700). G.C. acknowledges support from the National Key Research and Development Program of China (2023YFA1406101). Q.H. acknowledges support from China Postdoctoral Science Foundation (No. GZB20230421).

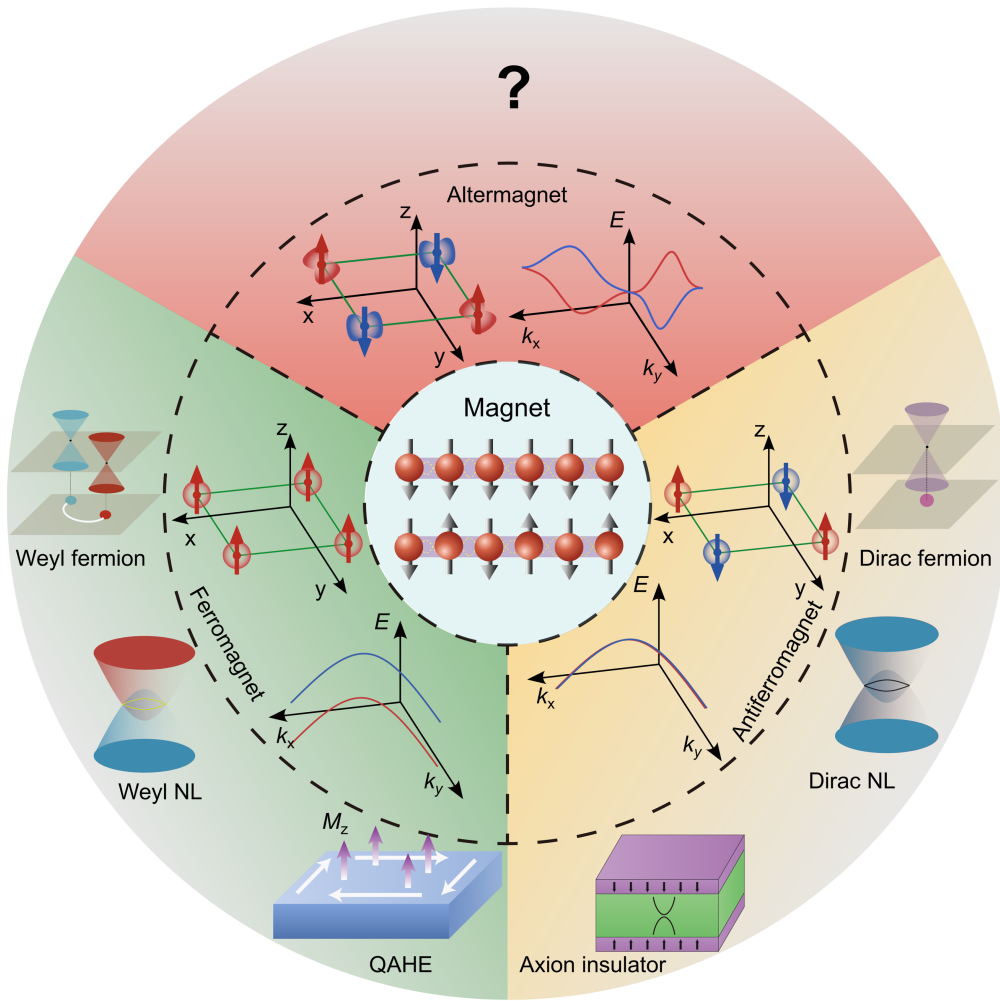

FIG. 1. The interplay between magnetism and other degrees of freedom, such as orbital or momentum, can give rise to a variety of novel quantum phenomena, including but not limited to quantized anomalous Hall effect [1], ferromagnetic Weyl semimetals [2,3]and antiferromagnetic axion insulators. In altermagnets, crystal symmetry connects antiferromagnetic sublattices in real-space while generating momentum-locked spin texture in reciprocal-space. This dual nature not only introduces spin and valley as tunable degree of freedoms for exploring novel quantum phases, but also establishes altermagnets as promising candidate for the next generation of electronic and spintronic device. Such a unique symmetry-governed interplay makes this frontier an exciting and largely unexplored territory in condensed matter physics.

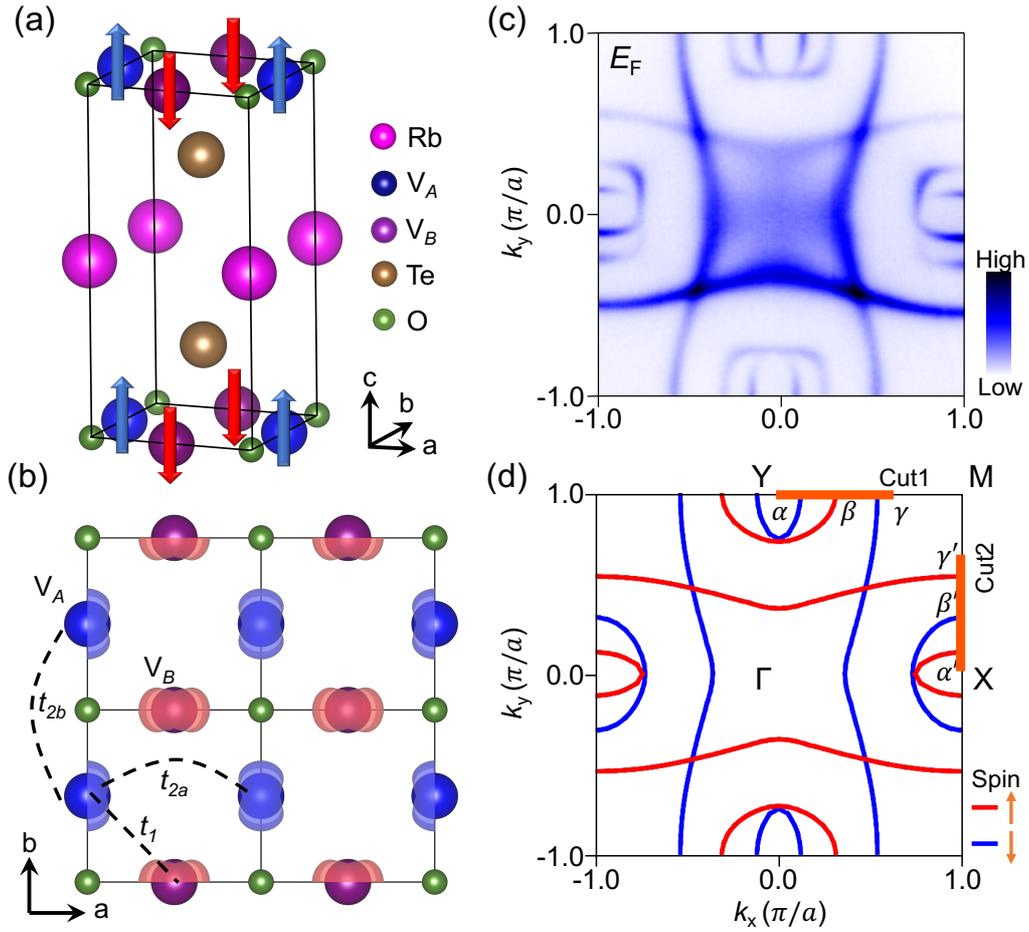

FIG. 2. Crystal structure and Fermi surface. (a) Schematic illustration of the crystal structure and magnetic structure of RbV$_2$Te$_2$O, where arrows indicate the magnetic moments of V which are parallel or antiparallel to the c axis. (b) Spin density of RbV$_2$Te$_2$O, which can be viewed as a variant of the Lieb lattice. The two magnetic sublattices, V$_A$ and V$_B$, are related by four-fold rotational symmetry ($C_4$) and mirror ($M_{xy}, M_{\bar{x}y}$). Due to the presence of the nonmagnetic O site, V$_A$ and V$_B$ sublattices experience distinct local environments, which give rise to anisotropic second nearest-neighbor hopping parameters, $t_{2a}$, $t_{2b}$. (c) Fermi surface map of RbV$_2$Te$_2$O measured by ARPES at $h\nu$ = 70 eV and $T$ = 20 K. (d) Spin-resolved DFT calculations of the Fermi surfaces at the $k_z = 0$ plane, with red and blue lines denoting spin-up and spin-down, respectively.

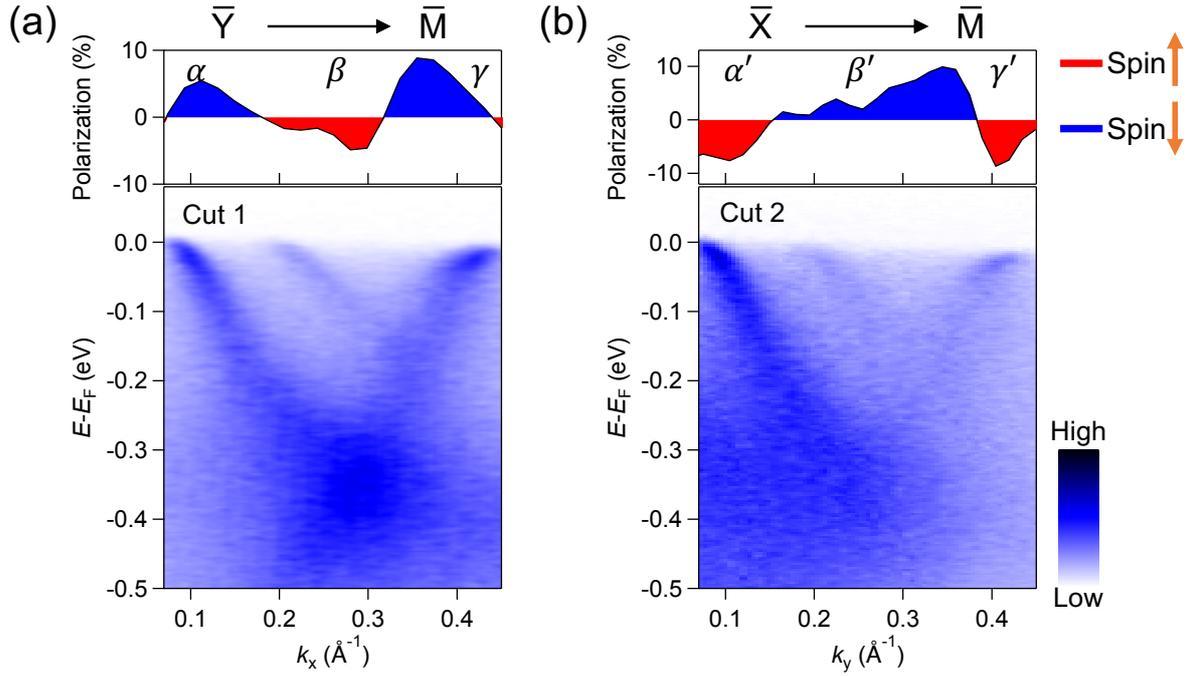

FIG. 3. Spin-resolved ARPES measurement. (a), (b) ARPES intensity plots along cut 1 (a) and cut 2 (b) measured at $hv$ = 70 eV and $T$ = 20 K. The ARPES intensity plot are the sum of the data collected under left-hand circularly polarization (CL) and right-hand circularly polarization (CR). The upper panels show the momentum-dependent spin polarizations calculated by the asymmetry of the spin-up and spin-down signals. Red and blue filled areas highlight the spin-up and spin-down polarizations. The spin-resolved ARPES data were collected at 13 K with $hv$ = 23 eV under LH polarization.

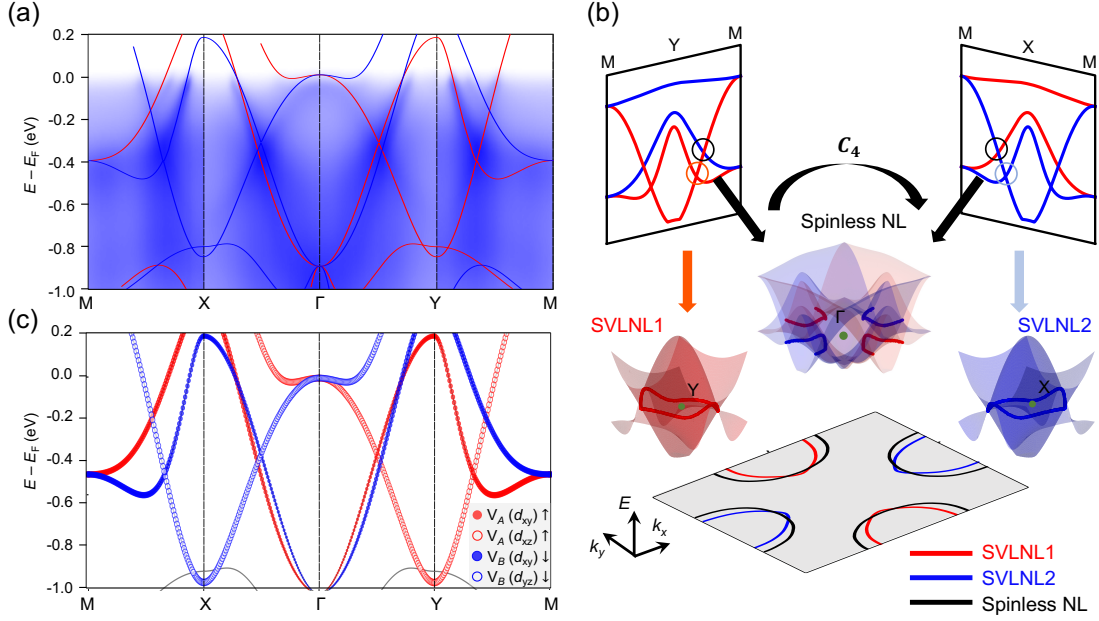

FIG. 4. Two types of nodal lines in RbV$_2$Te$_2$O. (a) ARPES intensity plot measured at 20 K with $h\nu$ = 70 eV. Overlaid solid curves represent DFT-calculated band structures, with red and blue indicating spin-up and spin-down states, respectively. (b) The momentum-dependent spin polarization leads to multiple twofold degeneracies on the mirror planes $k_z = 0$, giving rise to two distinct types of nodal lines: spin-valley locked nodal lines (SVLNLs) formed by crossings of same-spin bands with opposite spin polarizations in different valleys and spinless nodal lines (Spinless NL) formed by opposite-spin crossings. (c) Calculated orbital-resolved band structure along high-symmetry lines. Bands around the fermi level predominantly originate from the $d_{xz}^{\uparrow}(M_z = -1)$ and $d_{xy}^{\uparrow}(M_z = +1)$ orbits of the $V_A$ atom, $d_{yz}^{\downarrow}(M_z = -1)$ and $d_{xy}^{\downarrow}(M_z = +1)$ from the $V_B$. Here, states with $M_z = +1(-1)$ are represented by solid (dashed) circles.

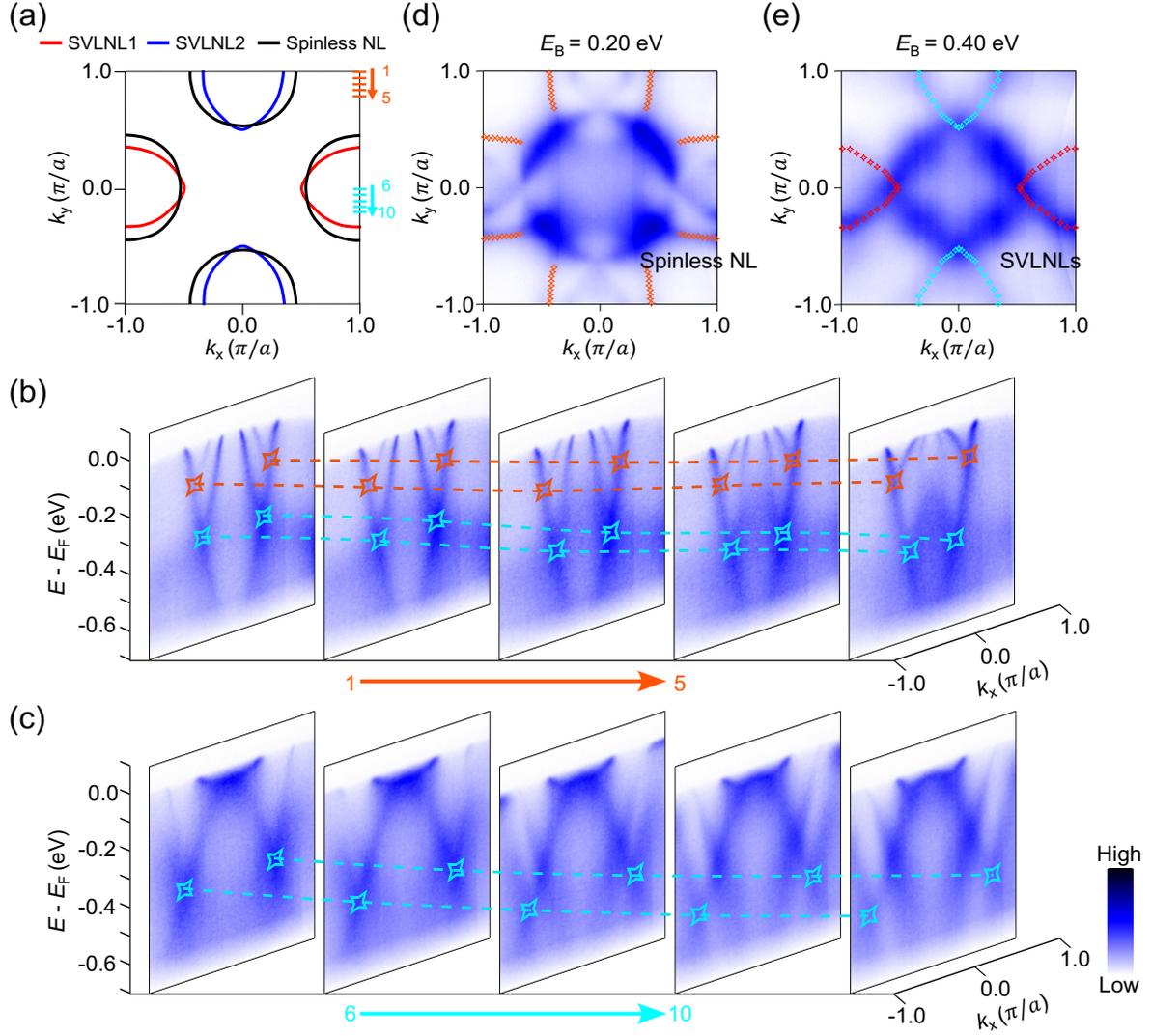

FIG. 5. Evidence for nodal lines. (a) Theoretically calculated two types of nodal lines in RbV$_2$Te$_2$O. (b), (c) Series of ARPES intensity plots along $k_x$ direction through the candidate node lines. The corresponding momentum locations of the intensity plots are indicated by cyan and orange lines in (a). The band crossing points, marked by cyan and orange four-pointed stars, are fitted with multiple Lorentzian peaks. (d), (e) Constant-energy surfaces of RbV$_2$Te$_2$O measured by ARPES at $hv$ = 70 eV and $T$ = 20 K, shown at binding energies $E_B$ = 0.2 eV and 0.4 eV. The momentum-space distribution of the experimentally observed nodal crossings is marked across the Brillouin zone.

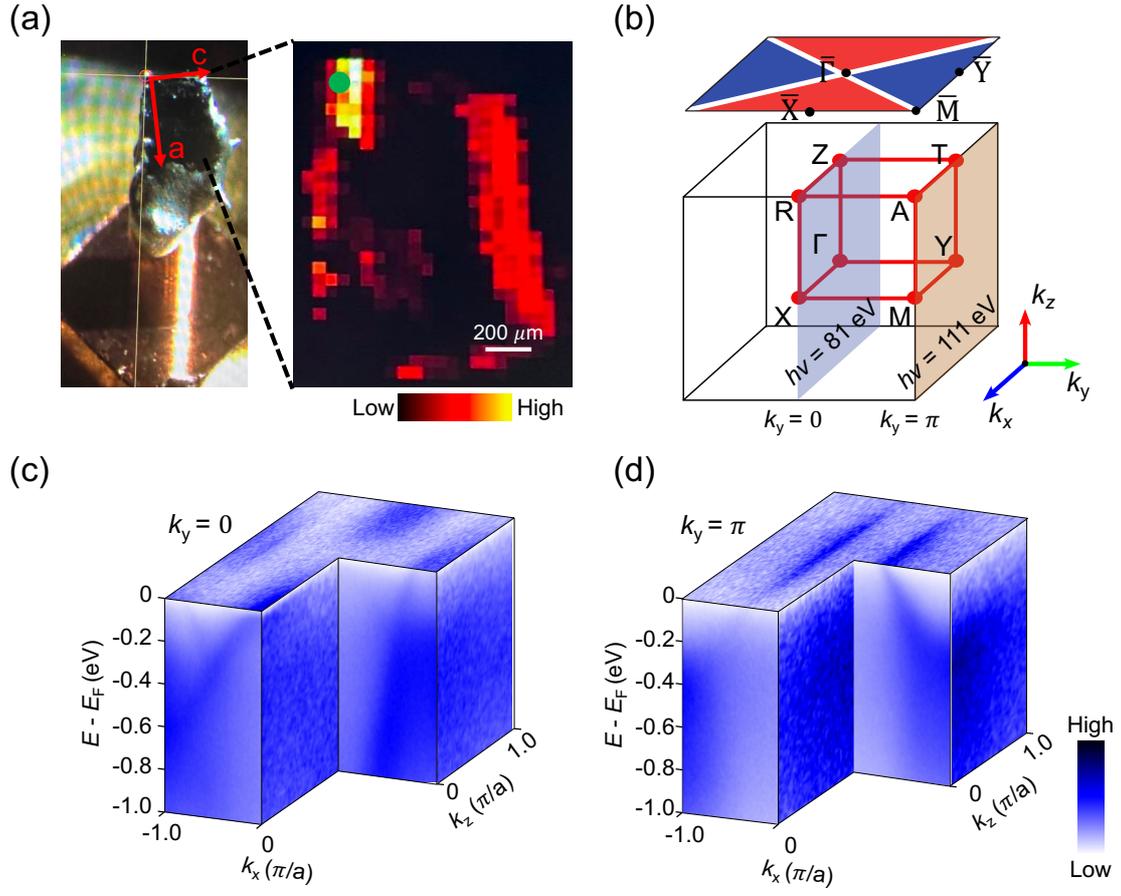

FIG. 6. Side-surface cleaving of RbV$_2$Te$_2$O. (a) Picture of Side-surface cleaved RbV$_2$Te$_2$O and real-space photoemission intensity mapping from a synchrotron-based ARPES. (b) Schematic of the bulk Brillouin zone (BZ) of RbV$_2$Te$_2$O. Photon energies of $hv$ = 81 eV and $hv$ = 111 eV correspond approximately to the $k_y$ = 0 and $k_y$ = π planes, respectively. (c), (d) Three-dimensional view of the electronic structure of RbV$_2$Te$_2$O measured at the $k_y$ = 0 and $k_y$ = π planes with linear horizontal (LH) polarization.

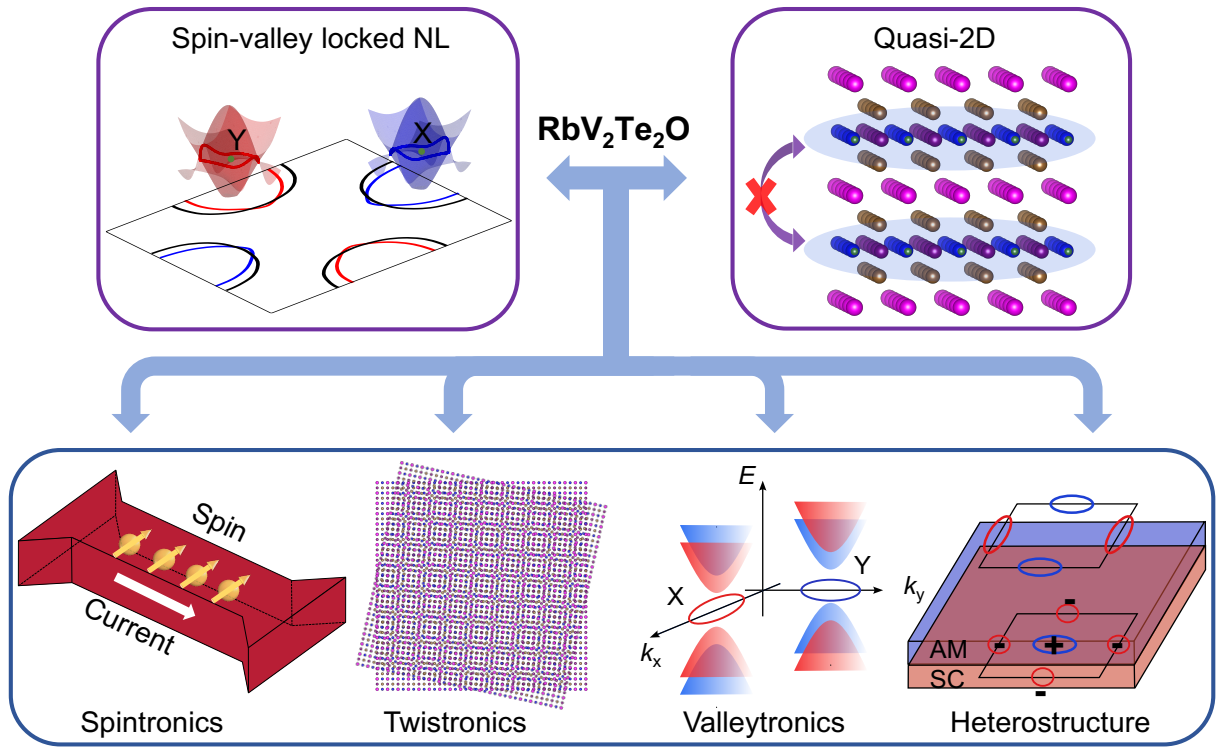

FIG. 7. RbV$_2$Te$_2$O is identified as a quasi-two-dimensional altermagnet hosting a previously unobserved class of spin–valley–locked nodal lines. The combination of its quasi-two-dimensional nature and topological characteristics opens the door to artificial structuring—including spintronics, twist-angle modulation, valleytronics and heterostructure engineering—providing a versatile platform for the realization and control of exotic quantum phases governed by the interplay of spin, valley, superconductivity and topology.

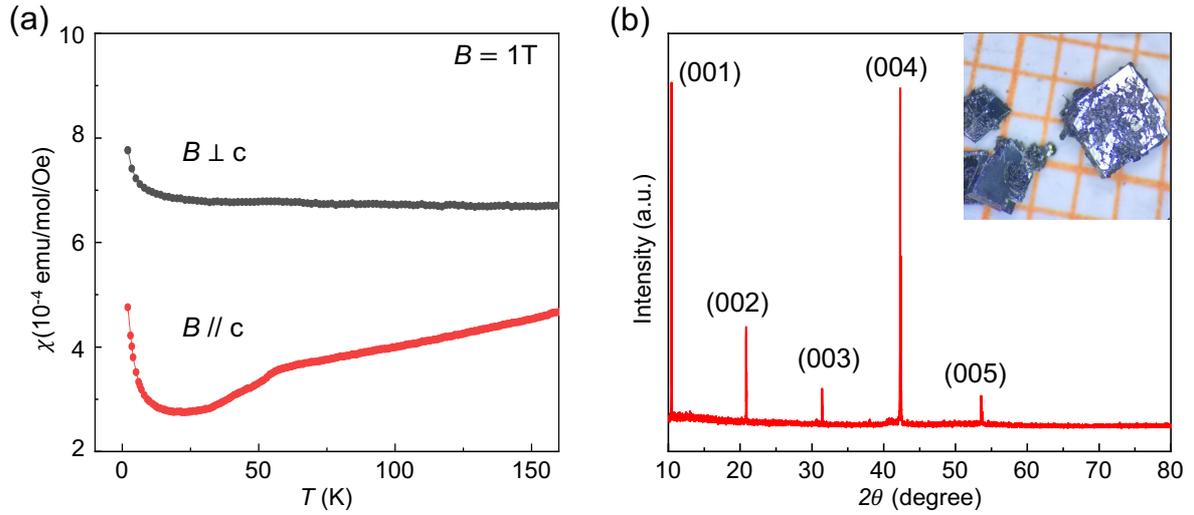

FIG. 8. Sample Characterization. (a) Temperature dependence of magnetic susceptibility $\chi$ measured under a magnetic field of 1T applied parallel to the $a$ and $c$ axis. (b) XRD pattern of a single crystal taken at 300 K. The inset shows a typical single crystal of $RbV_2Te_2O$.

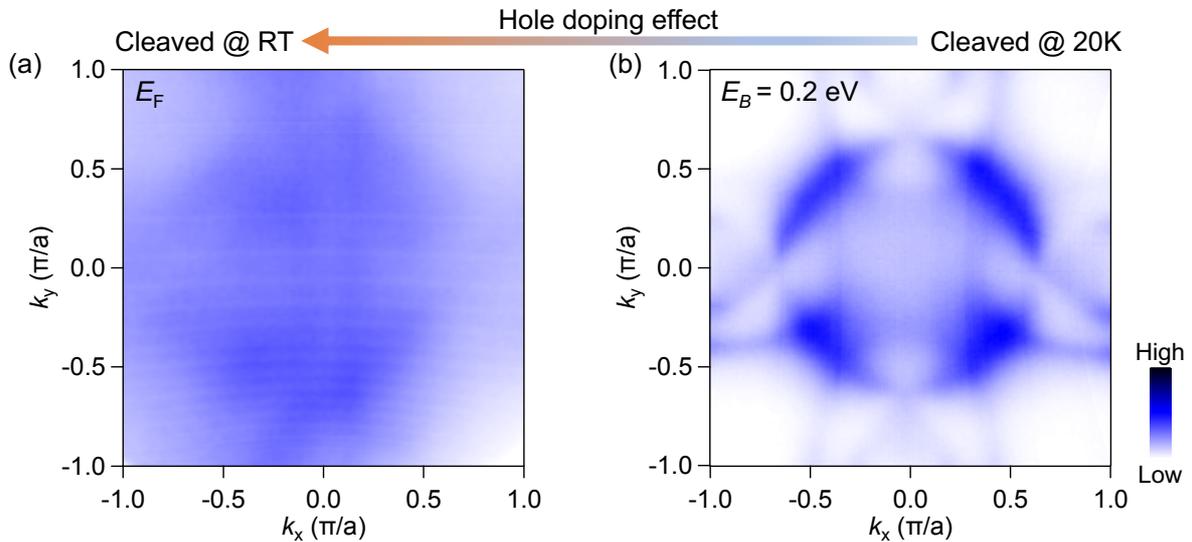

FIG. 9. Temperature-dependent cleavage–induced doping effect in $RbV_2Te_2O$. (a) Fermi surface measured on a sample cleaved at room temperature. (b) Constant-energy surface at a binding energy $E_B = 0.2$ eV obtained from a sample cleaved at 20 K. The two measurements exhibit a remarkable similarity. The enhanced desorption of surface Rb atoms during room-temperature cleavage leads to an effective hole doping effect to the electronic states.

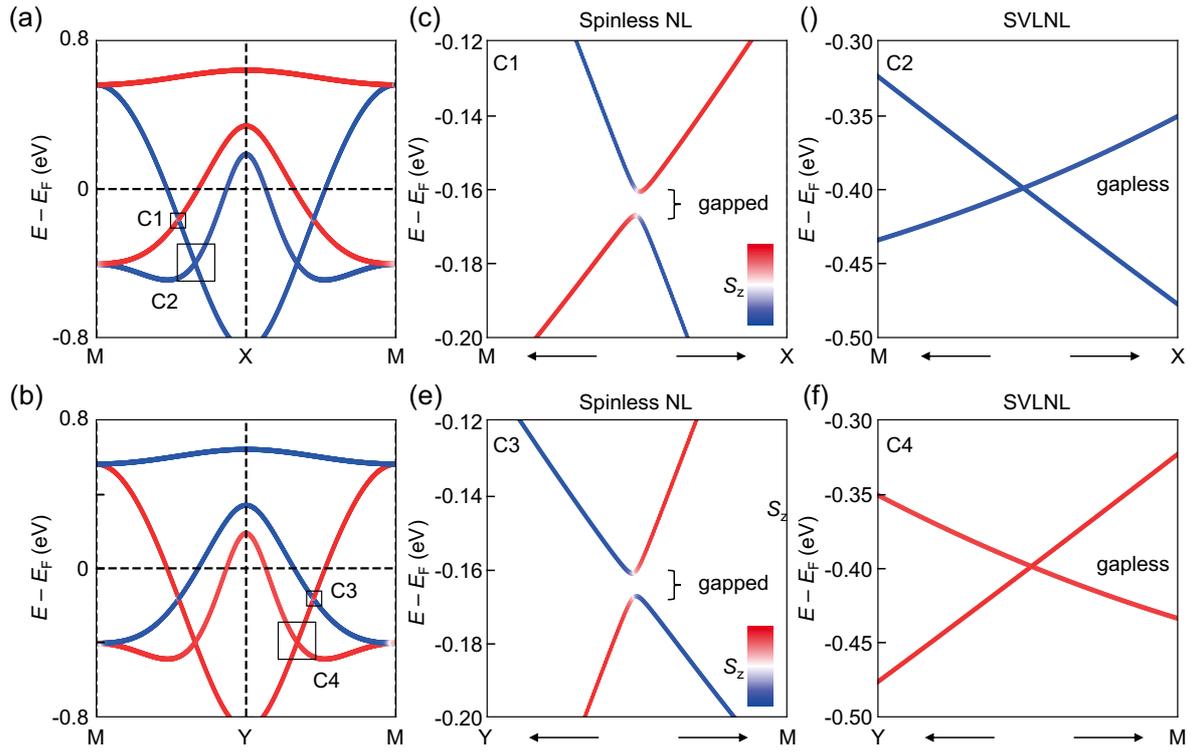

FIG. 10. DFT calculated band structure with spin-orbital coupling. (a), (b) Calculated electronic band structures along high-symmetry lines with SOC. (c), (d) and (e), (f) Zoomed views of the regions marked by boxes in (a) and (b), highlighting the fine band structures. The red and blue colors denote bands with opposite spin polarizations.

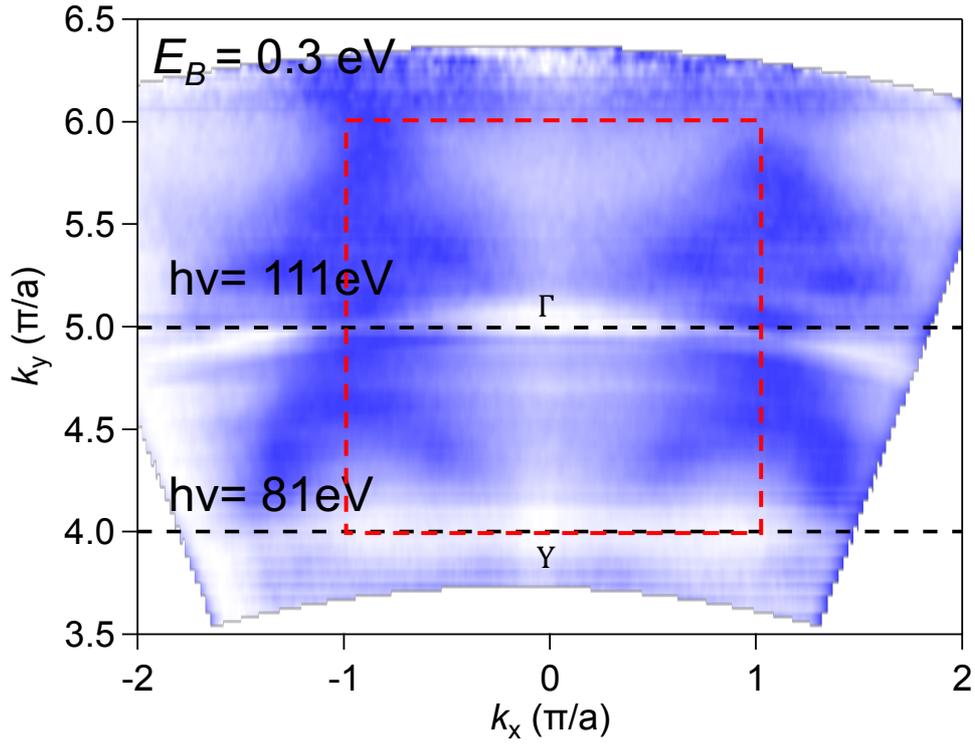

FIG. 11. ARPES constant-energy map collected in the x-y plane with $E_B$ =0.3 eV in a range of photon energies from 60 to 150 eV.

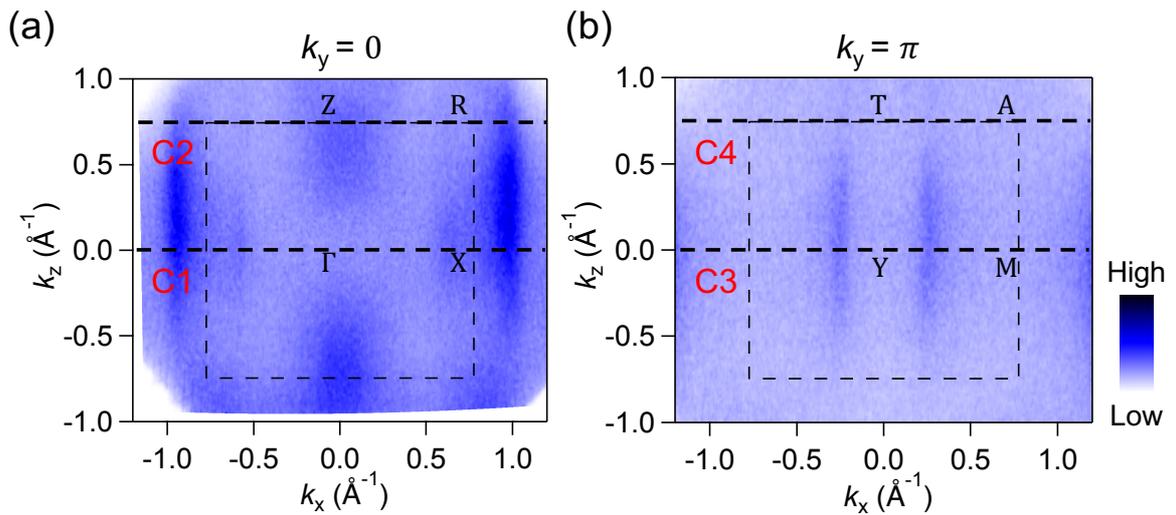

FIG. 12. Fermi surface measured from the x-z plane. (a), (b) Fermi surface maps measured by ARPES at $h\nu$ = 81 eV and 111 eV, respectively.

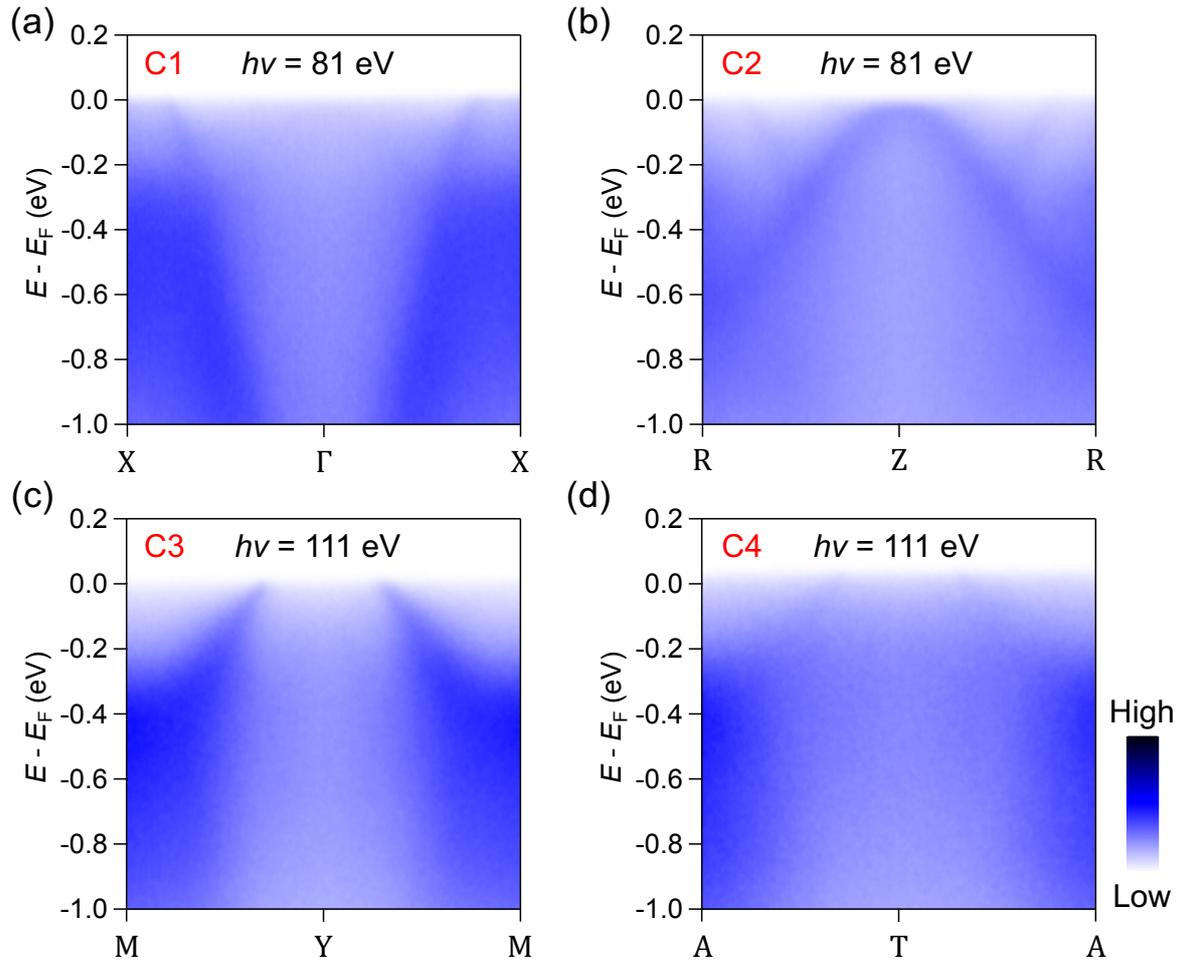

FIG. 13. Band structure of the side *x-z* surface. (a), (b) Band structure measured with 81 eV photons along Cut 1-2. (c), (d) Band structure measured with 111 eV photons along Cut 3-4. The band structure along high-symmetry directions exhibit negligible $k_z$ dispersion.